- RESEARCH ARTICAL -

# Performance Evaluation of Mobile Base Station under different Network Sizes on Cluster-Based Wireless Sensor Networks

Kadir Tohma[1*], İpek Abasıkeleş-Turgut[2], Cuma Celal Korkmaz[3], Yakup Kutlu[4]

[1,2,4] Department Of Computer Engineering, Faculty Of Engineering And Natural Sciences, Iskenderun Technical University, 31220 Iskenderun, Hatay, Turkey
[3] Department Of Electrical and Electronics Engineering, Faculty Of Engineering And Natural Sciences, Iskenderun Technical University, 31220 Iskenderun, Hatay, Turkey

**Abstract**
The position of the base station (BS) in wireless sensor networks (WSNs) has a significant impact on network lifetime. This paper suggests a mobile BS positioning algorithm for cluster-based WSNs, which considers both the location and the remaining energy level of the cluster heads in the network and evaluate the performance of the algorithm under different values of network sizes, including 100m x 100m, 200m x 200m and 300m x 300m. Simulations are conducted by using OMNeT++ and proposed method is compared with two different static BS positions, including central and external, on HEED protocol. The results show that mobile BS performs better than both central and external BS positions under all network sizes. Besides, the performance difference between the proposed method and the others increases as the size of the network increases, which demonstrates that the proposed mobile BS positioning also provides scalability.

## Introduction

The energy resources of WSNs are limited and often it is not possible to replace or renew them. Therefore, an energy efficient routing in these networks is quite important (Tohma, K *et al.*, 2015; Tohma, K *et al.*, in press). Routing in WSNs can be categorised as flat, hierarchical and adaptive. Hierarchical routing protocols, including LEACH (Heinzelman, W *et al.,* 2000) and HEED (Younis, O., & Fahmy, S., 2004), hold upper hand against its alternatives due to its energy efficiency.

---

[*] *Corresponding Author: Kadir Tohma, email: kadir.tohma@iste.edu.tr*

The location of BS plays an important role on decreasing the energy consumption of the nodes and accordingly increasing the lifetime of the network in hierarchical routing protocols (Abasıkeleş-Turgut, İ., 2016). Since, if the BS is far away from the network, the nodes have to consume much more energy to transmit their data to BS. Hence, their batteries drains away sooner and correspondingly the lifetime of the network is badly affected. The positions of BS is investigated in two categories in literature, including the static and dynamic (mobile) positioning. In a typical WSN topology, a single BS performs the duty of data collection on a fixed position. This position is only determined during the initial setup, is not changed along the lifetime of the network and is generally defined as the geometric centre of the gravity of the network (Luo, J., & Hubaux, J. P., 2005). On the other hand, the position of BS in mobile approach, frequently changes from the initial setup until the accomplishment of the task in such a way that it provides optimum dynamic communication performance (Cayirpunar, O *et al.,* 2013).

This paper investigates the effect of mobile BS under different networks sizes and compares the proposed mobile BS positioning method with two different static BS positions, including central and external, on HEED (Younis, O., & Fahmy, S., 2004), which is a popular hierarchical routing protocol in literature, by conducting various simulations.

The rest of this paper is organized as follows. Literature review is presented in Section II, whereas the structure of HEED algorithm, the proposed mobile BS positioning algorithm, simulation framework and parameters are described in Section III. Section IV includes the simulation results and discussion while Section V concludes the paper.

Abasıkeleş-Turgut (2016) has investigated the effect of four different BS locations on the performance of distributed and centralized cluster-based WSNs under different network sizes and the results show that the location of BS is quite determinant on the lifetime of the network. Since the location of the BS plays an important role on the performance of WSNs, a plenty of studies focuses on designing mobile BS positioning algorithms in literature.

Mollanejad et. al. (2010) have proposed a mobile BS positioning algorithm based on genetic algorithm and evaluated the performance of their algorithm on LEACH (Heinzelman, W *et al.,* 2000) and HEED (Younis, O., & Fahmy, S., 2004) protocols. Liang et al. (2010) has argued that the mobile BS movement should be limited. Therefore, they have suggested an algorithm for the optimum tour of BS with local movement. They have compared their algorithm with a problem, which they have programmed in a linear fashion, and stated that the proposed algorithm is more advantageous in terms of network lifetime. Salim and Badran (2015) have performed the movement of the mobile BS in accordance with several mobility scenarios presented in literature, including random, circular, middle way and rectangular. These scenarios have been investigated on various hierarchical routing protocols, including PEGASIS, PEGASIS-E, IECBSN, ECBSN, COSEN, EAPHRN. It has been reported that whole BS mobility scenarios, independently from the utilized protocol, provide superior results in comparison to static BS positioning. Wu and Chen (2007) have used dual BS: one of the base stations is static, while the other one is mobile. They restrict the movement of mobile BS by relocating the mobile BS only once in accordance with the location information of the nodes. They have observed that the usage of dual BS performs better than both single static and single mobile BS positions.

## Material and Method

### *The structure of HEED Protocol*

As is seen in Figure 1, the nodes are organized as clusters in hierarchical routing protocols. Each cluster has a cluster head, which collects data from the member nodes of its cluster, performs data aggregation and send data to BS. Member nodes of clusters are responsible from sensing and sending their data to cluster head.

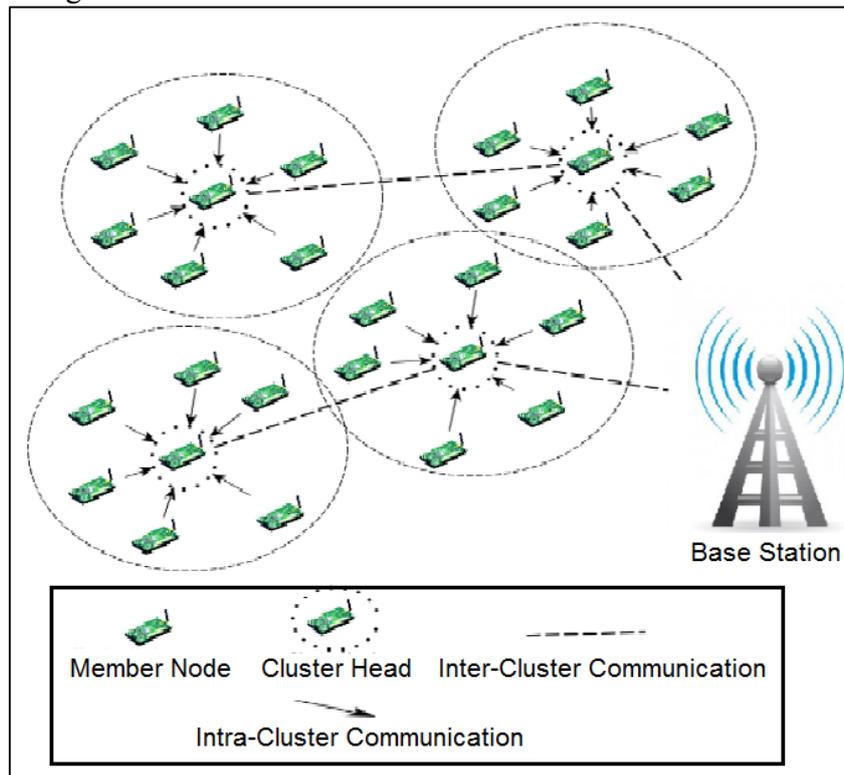

**Figure 1**. General structure of hierarchical routing protocols

HEED (Younis, O., & Fahmy, S., 2004), which is a popular hierarchical routing protocol, uses a series of iterations for carrying out the formation of clusters and deciding the cluster heads by using the remaining energy level of the nodes as a primary parameter and the distance of the nodes to each other as a secondary parameter (Equation 1). In the equation, $E_{residual}$ represents the remaining energy of the node, while $E_{max}$ states the initial energy level. $C_{prob}$ is the initial percentage of desired cluster heads in the network and $CH_{prob}$ represents the probability of a node to become a cluster head for that round.

$$CH_{prob} = C_{prob} * \frac{E_{residual}}{E_{max}} \quad (1)$$

Since the clustering procedure is repeated iteratively in short periods, cluster heads frequently changes when the algorithm finds a better candidate, which uniformly distributes the energy consumption over the network and has relatively higher average residual energy compared to the member nodes. The flowchart of the HEED can be seen in Figure 2.

### *The Proposed Mobile BS Positioning Methodology*

In this paper, the proposed mobile BS positioning algorithm, which is evaluated on HEED protocol by using OMNeT++ simulation program (Varga, A., 2001), is based on two parameters. One of them is the remaining energy level of the cluster heads, while the other is the location of the cluster heads. Since the inter-cluster data communication occurs between the cluster heads and the BS, in this paper, the movement of mobile BS is determined according to the cluster heads for energy efficiency, differently from the studies literature.

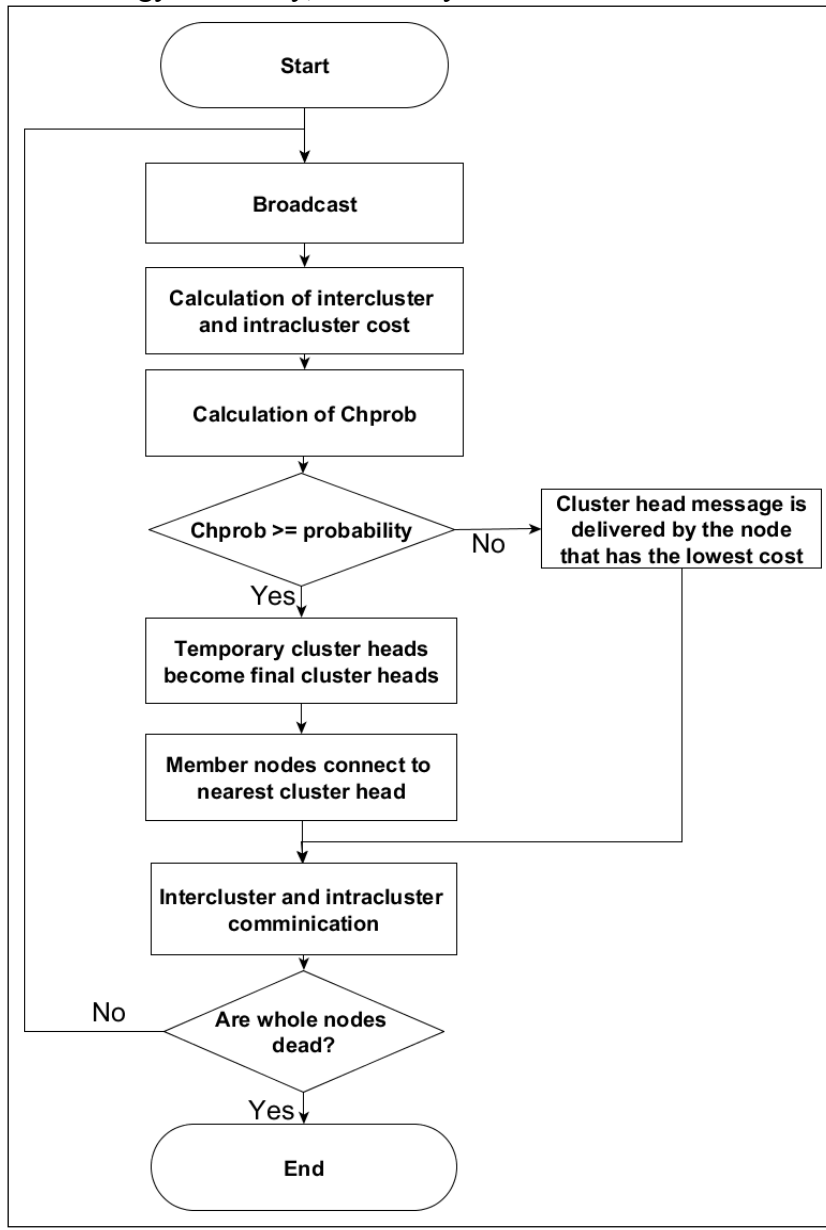

**Figure 2**. The flow chart of the HEED protocol

The remaining energy level of the cluster heads is chosen as a parameter because the lifetime of the network is directly related to the batteries of the nodes. In addition to this, the location of the cluster heads also play an important role in network lifetime since the energy consumption during the inter-cluster communication is based on this parameter. Therefore, the proposed BS mobility algorithm, called the location weighted with battery (LWB), is based on a formula, as is seen in Equation 2 and Equation 3, which uses the location of the cluster heads weighted with their remaining energy level. In each round, the BS is moved once according to these formulas.

$$LWBx = \left(\sum_{i=1}^{N} x_i E_r\right)/E_T \quad (2)$$

$$LWBy = \left(\sum_{i=1}^{N} y_i E_r\right)/E_T \quad (3)$$

In the equations, $LWB$ represents the location weighted with battery; $x_i$ ($y_i$) states the x (y) coordinate of the i.th node; $E_r$ is the residual energy of the node, while $E_T$ is the total initial batteries of the nodes.

The nodes consume energy proportional to the distance between the BS and the distance to each other. Therefore, the location of the BS has been frequently chosen as the geometrical centre of the network in literature. In this paper, the proposed algorithm is compared with two different static BS locations, including the central and external BS. In central BS approach, the BS is located in the geometrical centre of the gravity of the network, as used in literature. In external BS approach, the BS is located outside the network area, as in original HEED protocol.

*Simulation Framework and Parameters*

The proposed mobile BS positioning approach with central and external BS placement strategies are simulated on HEED protocol by using OMNeT++ (Varga, A., 2001), which is a modular network simulator, for different values of network sizes. The parameters used in simulations can be seen in Table 1.

Table 1. Simulation Parameters

| Parametres | Values |
|---|---|
| Network Area | 100m x 100m, 200m x 200m, 300m x 300m |
| The Number of Nodes | 100 |
| Probability of Cluster Heads | 0.05 |
| Energy consumed in the electronics circuit to transmit or receive the signal ($E_{elec}$) | 50 nJ/bit |
| Energy consumed by the amplifier to transmit at a short distance ($E_{fs}$) | 10 pJ/bit/m2 |
| Energy consumed by the amplifier to transmit at a longer distance ($E_{mp}$) | 0.0013 pJ/bit/m4 |
| Data Aggregation Energy ($E_{DA}$)($E_{fusion}$) | 5 nJ/bit/signal |
| Data Packet Size | 500 bytes |
| Broadcast Packet Size | 25 bytes |
| Packet Header Size | 25 bytes |
| Cluster radius | 25 m |
| Initial Energy ($E_0$) | 0.25 J |
| Threshold distance ($d_0$) | 75 m |
| Central BS Positions | (50,50) , (100,100) and (150,150) |
| External BS Positions | (50,175) , (100,350) and (150,525) |

**Results and Discussion**

For investigating the performance of both mobile and static BS positions, the number of simulation rounds when a certain number of nodes, including 1, 10, 20, 30, 40, 50, 60, 70, 80, 90, 95 and 100, are dead in the system for different sizes of networks is estimated, as is seen in Figure 3 through Figure 5. The lifetime of the network signifies the round, which all of the nodes (100) are dead. The round of the death of the intermediate nodes show the quality of the network because later the death of the nodes implies the higher rates of sensing data.

As is seen in Figures, mobile BS performs the best along the lifetime of the network, i.e. the death of all numbers of nodes occurs at the highest round, regardless of the size of the network. Central BS is hard on the heels of mobile BS, while the performance of external BS is the worst by landslide. This results show that mobile BS positioning algorithm provides the nodes to use their energies efficiently by decreasing the inter-cluster communication cost. Therefore, the nodes can live longer and both the lifetime and the quality of the network increase.

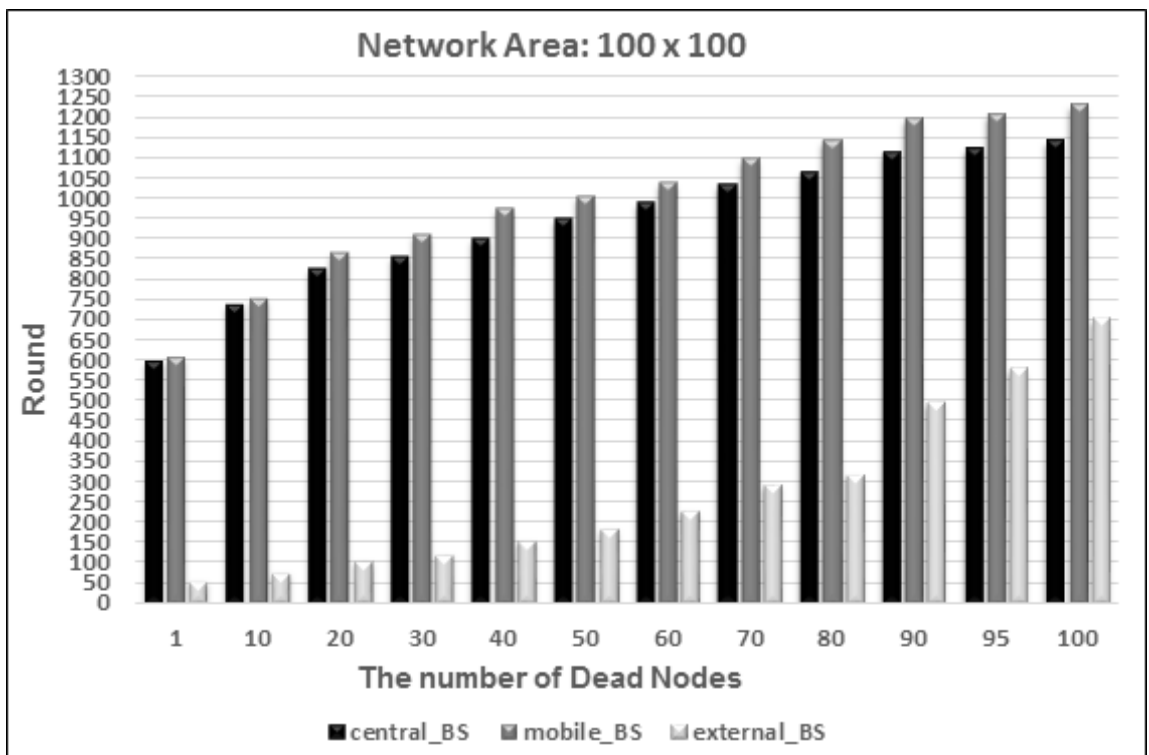

**Figure 3**. The number of rounds when a certain number of nodes are dead on 100m x 100m network.

The proposed mobile BS approach yields 5.7% higher performance on average and 8% higher performance at maximum than the central BS positioning approach on 100m x 100m network as is seen in Figure 3. When the networks size is enlarged to 200 m x 200 m (Figure 4), the rates goes up to 9.1% and 17.2%, respectively. Finally, when the network size becomes 300m x 300m (Figure 5), the highest performance increase is observed and the rates reaches up to 42% and 110%, respectively. The same tendency can also be seen between mobile and external BS locations. The difference of the performance between mobile BS and external BS starts from 400% (on a 100m x 100m network area) and comes to 6000% (on 300m x 300m network area) on average; 1001% and 10000% at maximum. The results show that the advantage of mobile BS positioning increases as the size of the network increases. The reason for this results underlies the alteration of the performance of the positioning algorithms as the network size increases.

As the size of the network increases, the nodes become dead at earlier rounds for all BS positioning approaches. The reason is that with the increasing size of the network, average distance between the nodes also increases. The more distance causes the more energy usage for data transmission during both inter-cluster and intra-cluster communication. The worst affected BS positioning approach from this situation is external BS. When the size of the network is upgraded from 100m x 100m to 200m x 200m, the decrease in the performance of external BS positioning goes up to 94%, while the network is enlarged from 200m x 200m to 300m x 300m, this rate arrives at 80%. Central BS is hard on the heels of external BS, while the performance of mobile BS is the least affected from the increase in the size of the network.

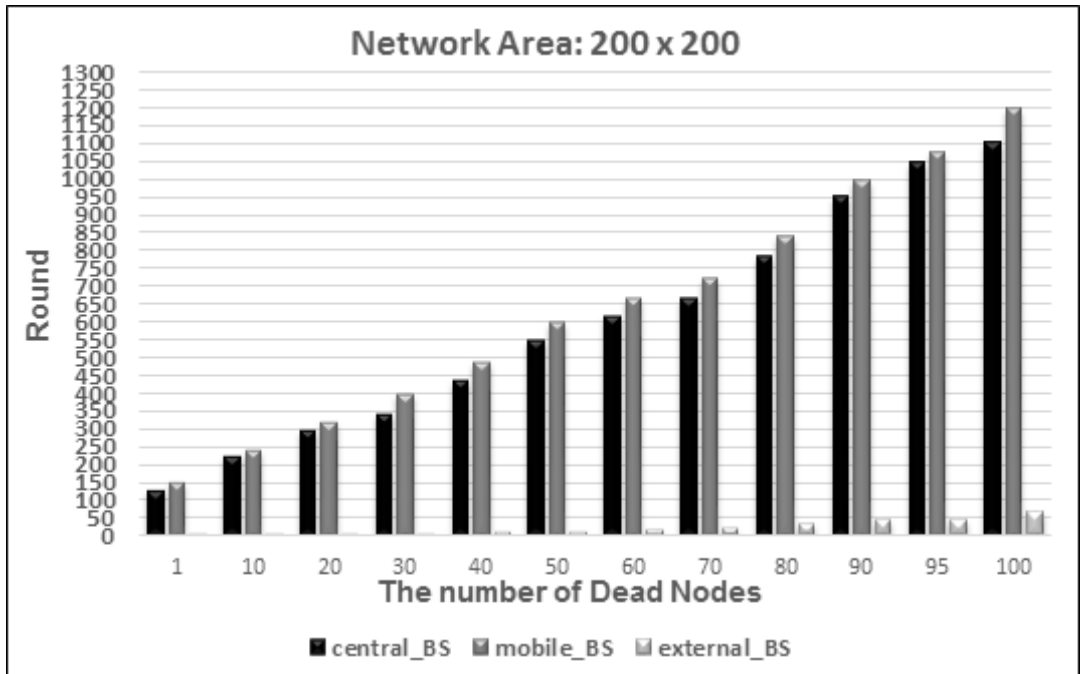

**Figure 4**. The number of rounds when a certain number of nodes are dead on 200m x 200m network.

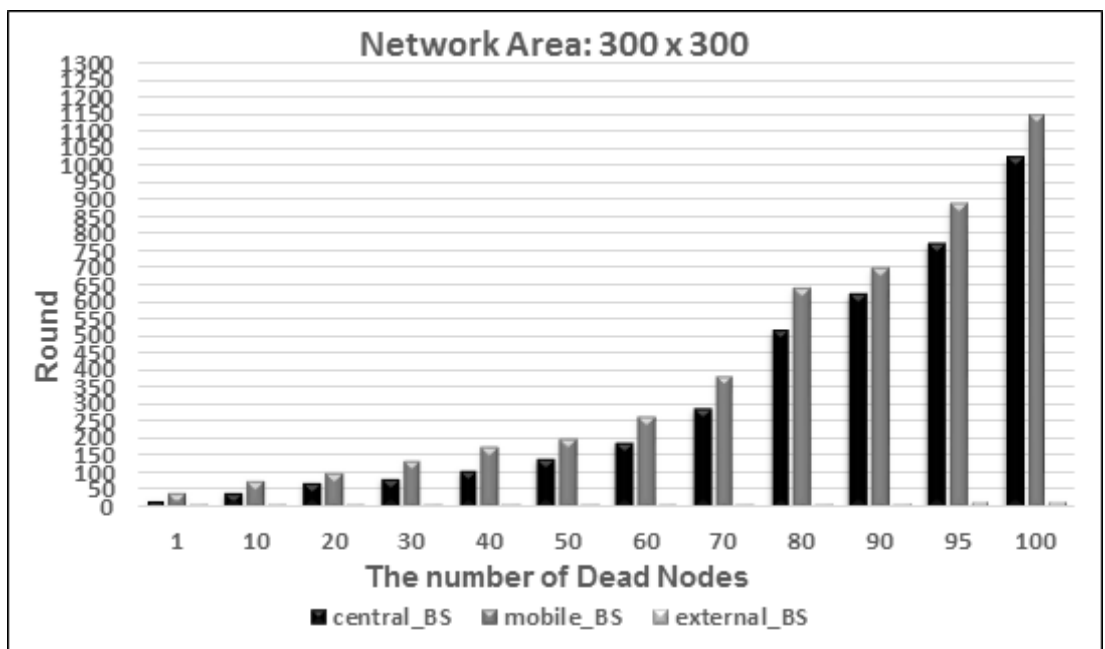

**Figure 5**. The number of rounds when a certain number of nodes are dead on 300m x 300m network.

Although the rates are close to each other, when the size of the network is widen from 100m x 100m to 200m x 200m, the decrease in the performance of central BS positioning goes up to 79%, while that of mobile BS positioning arrives at 75%. Accordingly, if the network is enlarged from 200m x 200m to 300m x 300m, this rates are 82% and 69% for central and mobile BS positioning, respectively. The optimum BS positioning approach, which decreases the communication cost between the nodes becomes much more important for larger networks, where the nodes consume extra energy for data transmission on every stages during the lifetime of the network due to the increased distance. Hence, the advantage of mobile BS positioning increases as the size of the network increases. This results show that another advantage of the proposed mobile BS positioning algorithm is scalability.

The position of the BS is a critical issue on the lifetime of the network in WSNs, in which the resources are limited, a plenty of studies in literature focuses on mobile BS positioning algorithms. Therefore, a mobile BS positioning algorithm that considers both the location and the remaining energy level of the cluster heads is suggested. The proposed algorithm is compared with two different static BS positioning approaches on HEED for different sizes of networks by using OMNeT++ simulation program. The results show that the proposed mobile BS positioning performs better in terms of the lifetime and the quality of the network than the static approaches for all sizes of networks. Additionally, as the size of the network increases, the performance of mobile BS draws away, which shows that the proposed algorithm also provides scalability.

**Acknowledgments**

This work has been supported by TUBITAK 115E211 EEEAG numbered project.